\documentclass[aps,twocolumn,superscriptaddress,nofootinbib]{revtex4-1}
\usepackage{amsmath}
\usepackage{amsfonts}
\usepackage{amssymb}
\usepackage{mathtools}
\usepackage{graphicx}
\usepackage[caption=false,position=t]{subfig}
\usepackage[colorlinks]{hyperref}
\usepackage[capitalize]{cleveref}
\usepackage{color}
\usepackage{siunitx}
\usepackage{upgreek}

\usepackage{yypreamble}
\newcommand{\subS}{\mathbb{S}}
\newcommand{\pardash}[1]{\paragraph*{#1 ---}}

\usepackage[normalem]{ulem}

\newcommand{\rev}[1]{{#1}}

\begin{document}

\title{Two-dimensional hard-core Bose-Hubbard model with superconducting qubits}

\author{Yariv Yanay}
\email{yariv@lps.umd.edu}
\affiliation{Laboratory for Physical Sciences, 8050 Greenmead Dr., College Park, MD 20740}
\author{Jochen Braum\"uller}
\author{Simon Gustavsson}
\affiliation{Research Laboratory of Electronics, Massachusetts Institute of Technology, Cambridge, USA, MA 02139}
\author{William D. Oliver}
\affiliation{Research Laboratory of Electronics, Massachusetts Institute of Technology, Cambridge, USA, MA 02139}
\affiliation{MIT Lincoln Laboratory, 2	44 Wood Street, Lexington, USA, MA 02421}
\affiliation{Department of Electrical Engineering \& Computer Science, Massachusetts Institute of Technology, Cambridge, USA, MA 02139}
\affiliation{Department of Physics, Massachusetts Institute of Technology, Cambridge, USA, MA 02139}
\author{Charles Tahan}
\affiliation{Laboratory for Physical Sciences, 8050 Greenmead Dr., College Park, MD 20740}

\date{\today}

\begin{abstract}
The pursuit of superconducting-based quantum computers has advanced the fabrication of and experimentation with custom lattices of qubits and resonators.
Here, we describe a roadmap to use present experimental capabilities to simulate an interacting many-body system of bosons and measure quantities that are exponentially difficult to calculate numerically.
We focus on the two-dimensional hard-core Bose-Hubbard model implemented as an array of floating transmon qubits. 
We describe a control scheme for such a lattice that can perform individual qubit readout and show how the scheme enables the preparation of a highly-excited many-body state, in contrast with atomic implementations restricted to the ground state or thermal equilibrium. 
We discuss what observables could be accessed and how they could be used to better understand the properties of many-body systems, including the observation of the transition of eigenstate entanglement entropy scaling from area-law behavior to volume-law behavior.
\end{abstract}

\maketitle

\section*{Introduction}

Analog quantum simulators have evolved in the last two decades from a theoretical concept to an experimental reality (see e.g.~\cite{Buluta2009,Cirac2012,Georgescu2014}). Initial experimental success was predominantly achieved with atomic systems, including neutral gases and trapped ions~\cite{Greiner2002,Friedenauer2008,Gerritsma2010,Schneider2012,Greif2013}. 
More recently, superconducting circuits have emerged as a viable quantum simulation platform \cite{Houck2012,Marcos2013,Schmidt2013,Devoret2013,Neill2018}. This modality -- based on ``artificial atoms'' -- features a high degree of experimental controllability and stability~\cite{Krantz2019}. 
The flexibility of the superconducting platform has enabled several successful quantum simulation experiments \cite{Roushan2017,Lamata2018,Kjaergaard2019,Ye2019,Arute2019,Chiaro2019}.

\begin{figure}[tbp] 
   \centering

   \hfill
   \subfloat[HCB Lattice]{\includegraphics[width=0.6\columnwidth]{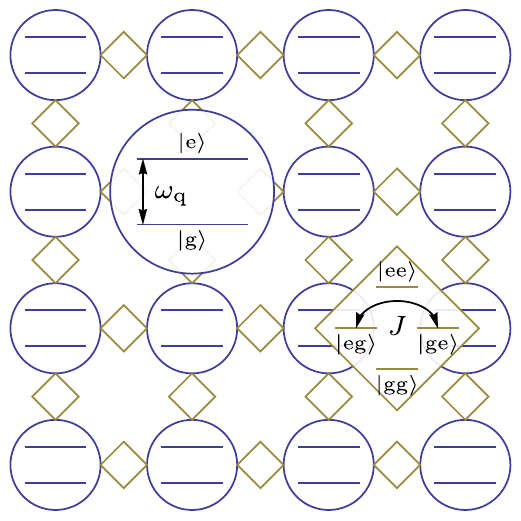} \label{fig:grid}}
   \hfill
   \subfloat[HCB Spectrum]{\includegraphics[height=0.6\columnwidth]{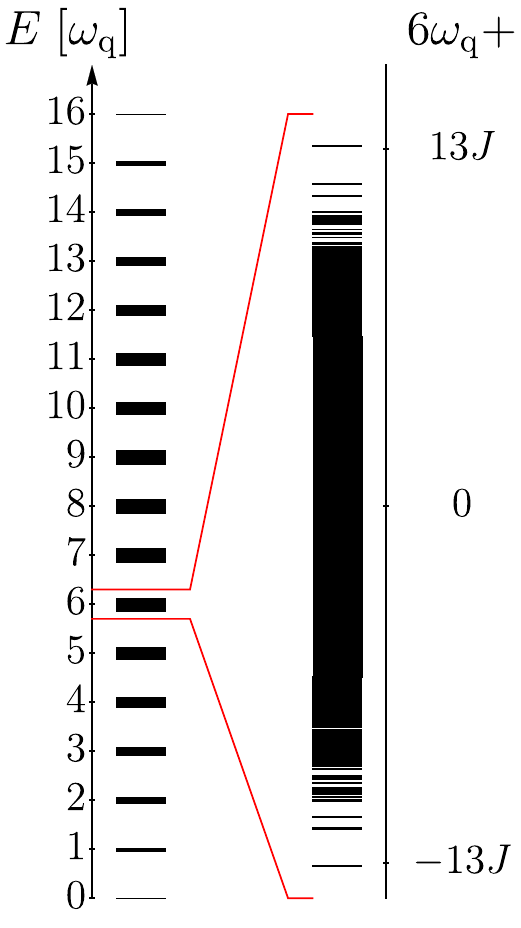} \label{fig:spectsketch}}
   \hfill\hfill

   \caption{\rev{{\bf The two-dimensional hard-core Bose-Hubbard model (HCB).}}
   \protect\subref{fig:grid}~Sktech of a sample $4\times4$ HCB lattice. Each circle represents a qubit, constrained to two energy levels with energy difference $\omega_{\rm q}$. The diamonds represent coupling between each pair of nearest neighbors at strength $J$. One magnified version of each element shows these energies.
	\protect\subref{fig:spectsketch}~The spectrum of the same system, here with $J=\omega_{\rm q}/10$. On the left we show the entire spectrum; for a system with $N$ qubits, it is composed of $N+1$ sectors defined by the total excitation number $n$. On the right, we show a close up of a particular sector, with an energy bandwidth $\Delta E \propto 8n\p{1-n/N}J$.
   }
   \label{fig:sketch}
\end{figure}

Here, we show how to realize the two-dimensional (2D) hard-core Bose-Hubbard model (HCB) illustrated in \cref{fig:sketch} using an array of transmon qubits \cite{Koch2007}, the current workhorse qubit design in superconducting circuits.
The HCB is a strongly interacting system that displays some of the critical properties of interacting quantum systems, including the area-law to volume-law transition of the entanglement spectrum that has been extensively studied in many-body systems \cite{Eisert2010}.
Outside of one dimension (1D), this system has no known analytical solution, and its study has been conducted mostly through numerical methods limited in their scope. The most successful approach has been the use of tensor network methods, which focus on finding the ground state energy \cite{Murg2007,Jordan2009}.
An experimental realization of a 2D HCB could offer new and complementary insights about the eigenstates and dynamics of many-body systems. It could also be used to validate the results of tensor network methods in large systems, and test their underlying assumptions on the nature of many-body wavefunctions. An experimental realization also offers access to the system's entire spectrum, allowing one to measure the many-body properties of its excited states. 

Previous experiments have realized the HCB in 1D \cite{Roushan2017,Ma2019}, where the model can be solved by analytical methods~\cite{Paredes2004,Girardeau1960} and has the dynamics of a free fermion gas. \rev{Recent realizations have also explored entanglement propagation in ladders and a $3\times7$ array \cite{Chiaro2019}.} Here, we propose the implementation of the 2D HCB with state-of-the-art transmon qubits. We calculate the requirements on qubit uniformity and lifetime, and describe the control systems required to measure the array's many-body properties. Finally, we propose a technique to generate highly-excited states that enable one to more completely explore the system's spectrum and observe its many-body properties.

\subsection*{Superconducting quantum many-body physics simulator}

We consider the implementation of a \rev{quantum many-body physics simulator (QMBS)} with a superconducting quantum circuit made up of multiple repetitions of small basic circuits implementing qubits\footnote{Note that while here and throughout the paper we take each site to be a qubit, i.e.~a two-level system, the discussion here applies equally to systems of spins or particles on a lattice, etc.} and coupling elements.
We describe the system with the Hamiltonian
\begin{equation}
\hat H = \sum_{i} \hat H^{\rm Q}_{i} +  \sum_{\avg{i,j}}\hat H^{\rm J}_{i,j},
\label{eq:HS}
\end{equation}
where summation is over all qubits $i$ and over all coupled pairs $\avg{i,j}$. The terms $\hat H^{\rm Q}_{i}$ and $\hat H^{\rm J}_{i,j}$ describe the basic qubit and coupling circuits, respectively.
The system in \cref{fig:sketch} is one example of the Hamiltonian of \cref{eq:HS}, with circles (qubits) representing $\hat H^{\rm Q}_{i}$ and diamonds (coupling elements) representing $\hat H^{\rm J}_{i}$.

\begin{table}[h]
   \centering
   \renewcommand{\arraystretch}{1.1}
   \begin{tabular}{|@{\hspace{0.5em}} ccl  @{\hspace{0.5em}}|@{\hspace{0.5em}} l |} 
   \hline \multicolumn{3}{|@{\hspace{0.5em}} l |@{\hspace{0.5em}}}{Energy scale} & Description 
\\   \hline
$\omega_{\rm q}$ & = &  $\overline{{}_{i}\bra{\rm e}\hat H\ket{\rm e}_{i}}$ &  Qubit frequency
\\ $A$  & = &   $ \overline{{}_{i}\bra{\rm f}\hat H\ket{\rm f}_{i}} -  2\omega_{\rm q} $  & Anharmonicity
\\ $J$ & = & $\overline{\overline{\abs{{}_{i}\bra{\rm e}\hat H\ket{\rm e}_{j}}}}$ & Hopping energy
\\ $\Delta\omega$ & = & $\sqrt{\;\overline{\p{{}_{i}\bra{\rm e}\hat H\ket{\rm e}_{i} - \omega_{\rm q}}^{2}}}$  & Frequency variance
\\   \hline
   \end{tabular}
   \caption{Energy scales of the QMBS.
   Here, $\ket{\rm e}_{i}$ ($\ket{\rm f}_{i}$) is the state with qubit $i$ in its first (second) excited state and all others in the ground state. $\overline{X_{i}}$ ($\overline{\overline{X_{i,j}}}$) denote the average of $X_{i}$ ($X_{i,j}$) over all qubits (all coupled pairs). We take $\hbar=1$ and the ground state energy to be zero. }
   \label{tab:Escales}
\end{table}

We note that the QMBS can be characterized by four energy scales derived from these Hamiltonians, outlined in \cref{tab:Escales}. 
The qubit frequency $\omega_{\rm q}$ and hopping strength $J$ are the typical energy scales of the qubit and coupling, respectively, and the anharmonicity $A$ describes the deviation of the qubits from harmonic level spacing. The frequency mismatch $\Delta\omega$ is the scale of non-uniformity across the system, including, e.g., variation introduced during fabrication.
We neglect deviations in the coupling strength, and assume that the deviations in the first level spacing are typical of the rest of the spectrum.

\begin{figure}[t] 
   \centering
   \includegraphics[width=\columnwidth]{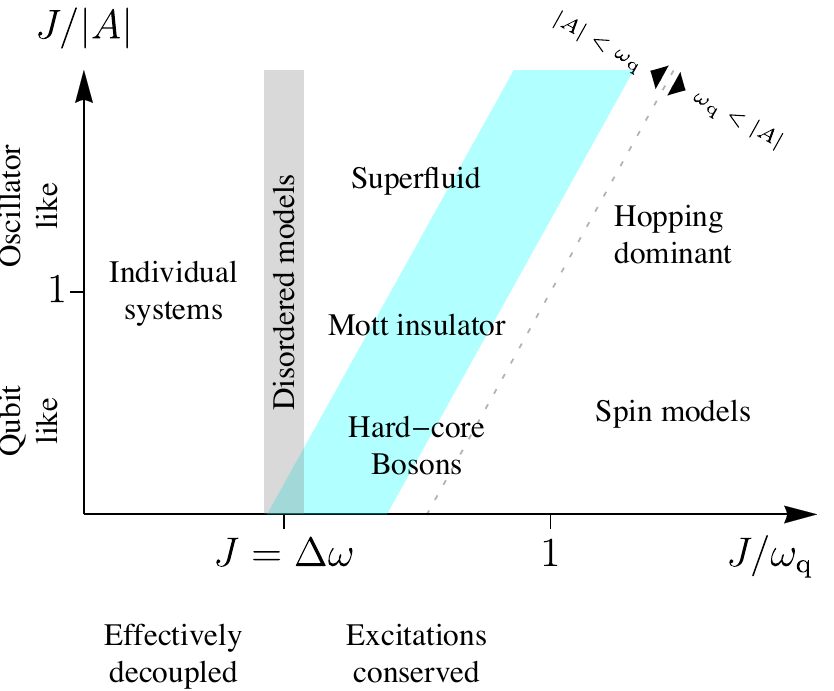} 
   \caption{\rev{\bf Accessible models with a QMBS based on a superconducting circuit.} Which models can be realized depending on the ratio of the coupling strength $J$ to the qubit frequency $\omega_{\rm q}$ and anharmonicity $A$. The diagonal cyan shading highlights the area accessible with transmon qubits, which are our focus here. 
   The coupling must be greater than the frequency spread, $J\gtrsim \Delta\omega$, for any kind of many-body physics to appear. 
   Where $\Delta\omega \lesssim J \ll \omega_{\rm q}$, the behavior is particle-like and we expect to see a version of the Bose-Hubbard model with $A$ playing the role of on-site interaction. 
   Where $\omega_{\rm q}\lesssim J \ll \abs{A}$ we have a spin-like model, where each unit acts as a two-level qubit while the coupling does not conserve excitation number. 
   When $J$ dominates all other scales, we expect semiclassical behavior. }
   \label{fig:phase}
\end{figure}

The behavior of the QMBS depends on the ratio of $J$ to the other three scales. At $J\ll \Delta\omega$, exchange of energy between different qubits is suppressed, and the system will behave as a collection of uncoupled circuits. In this case, there is no many-body physics to speak of, and the system decomposes into multiple systems with a single degree of freedom each. Thus $J\gtrsim \Delta\omega$ is required for many-body dynamics to appear in the lattice. The ratios $J/\abs{A}$ and $J/\omega_{\rm q}$ then determine which states are effectively coupled and thereby which theoretical models are accessible \cite{Sachdev2011}. These features are collected in the form of a phase diagram in \cref{fig:phase} and discussed in further detail below.

\pardash{Hopping dominant}

In the regime at the top right corner of \cref{fig:phase}, the dominant energy scale is $J$, the coupling energy. This describes systems such as quantum rotor models in the paramagnetic phase. We note that generally, for a large number of sites, these have a large density of states, and it may be difficult to prepare the system in a low-temperature quantum state. In that case, the system can be understood by a semiclassical description, and it is hard to observe uniquely quantum dynamics. Such experiments have been performed for large numbers of Josephson junctions \cite{vanderZant1996,Paramanandam2011}.

We note also that in other cases this regime can be avoided by choosing a different basis of states to describe the Hamiltonian, i.e.~by switching the choice of which circuits describe the qubits $\hat H^{\rm Q}_{i}$ and the coupling elements $\hat H^{\rm J}_{i,j}$.

\pardash{Particle-like models}
In the central portion of the phase diagram, the hierarchy of scales is
\begin{equation}
\Delta\omega \lesssim J \ll \omega_{\rm q}.
\end{equation}
Here, the rotating wave approximation is valid, and the coupling elements can move an excitation between sites but will not change the total number of excitations. This regime is equivalent to models of bosonic particles, and we may describe the system with the Bose-Hubbard Hamiltonian,
\begin{equation}\begin{gathered}
\hat H^{\rm Q}_{i} = \omega_{\rm q}\hat n_{i} + \half A \hat n_{i}\p{\hat n_{i} - 1},
\\\hat H^{\rm J}_{i,j} = - \rev{J_{xx}}\p{\hat a_{i}\dg\hat a_{j} + \hat a_{j}\dg\hat a_{i}} + \rev{J_{zz}}\hat n_{i}\hat n_{j},
\end{gathered}\end{equation}
where $\hat a_{i}\dg$ is the creation operator for site $i$ and $\hat n_{i} = \hat a_{i}\dg\hat a_{i}$ is its energy level. 
Here, the anharmonicity plays the role of the on-site interaction strength, while \rev{inter-qubit coupling generates transverse hopping terms \rev{($J_{xx}$)} and longitudinal interaction terms \rev{($J_{zz}$)}.}

The sub-regime where $J<\abs{A}\ll\omega_q$ -- the working point of the transmon qubit~\cite{Koch2007} -- is the most  experimentally accessible parameter regime and is widely adopted by the superconducting circuit community in a multitude of experiments (see e.g.~\cite{Kjaergaard2019}), including recent implementations of 1D Bose-Hubbard lattices~\cite{Ma2019,Yan2019}. In this manuscript, we focus on this regime.

We note that a subset of the particle-like regime, where $\Delta\omega \sim J$, can be used to simulate disordered systems. This can be achieved either by intentionally varying the qubit frequency across the lattice, or by decreasing the hopping energy at a constant residual disorder.

\pardash{Spin-like models}
At the bottom right corner of \cref{fig:phase}, the energy scales are given by
\begin{equation}
\Delta\omega \ll\omega_{\rm q}\lesssim  J \ll \abs{A}.
\end{equation}
Here, the anharmonicity dominates the coupling term, ensuring that each unit cell remains within the qubit manifold. However, the coupling elements are strong enough to change the qubit state in a non excitation-conserving way. The rotating wave approximation then breaks down, and the system is best understood by a spin-like model,
\begin{equation}\begin{gathered}
\hat H^{\rm Q}_{i} = \half[\omega_{\rm q}]\hat \gs^{z}_{i},
\qquad \hat H^{\rm J}_{i,j} = \sum_{\mu,\nu}J_{\mu,\nu} \hat\gs_{i}^{\mu}\hat \gs_{j}^{\nu}
\end{gathered}\end{equation}
where $\hat \gs^{\mu}_{i}$ are the Pauli operators on site $i$.

This regime, where the coupling strength becomes similar to the transition frequencies of the coupled systems, is known as the the ultra-strong or deep-strong coupling regime~\cite{Casanova2010,Forn-Diaz2019}, and it is more challenging to realize experimentally. However, superconducting artificial atoms are more suitable for its realization than natural atoms coupled to an electromagnetic cavity, as their coupling strength to a harmonic oscillator mode is not necessarily limited by the fine structure constant~\cite{Devoret2007,Manucharyan2017}. In general, physical couplings in the deep-strong coupling regime can be achieved with strongly non-linear qubits and high-impedance circuits~\cite{Manucharyan2017}. 
A promising qubit modality to reach such high couplings is the flux qubit, where $\omega_{\rm q}\ll \abs{A}$, as demonstrated experimentally \cite{Yoshihara2017,Forn-Diaz2017}. The fluxonium qubit \cite{Manucharyan2009}, an extension of the flux qubit, has recently been demonstrated to preserve long coherence times while in the high anharmonicity regime \cite{Nguyen2018}.

\section*{Results}

\subsection*{The Hard-Core Bose-Hubbard model}

For the remainder of this article, we focus our attention on the regime
\begin{equation}
\Delta\omega \lesssim J \ll \omega_{\rm q},\abs{A}.
\end{equation}
This combines the two constraints mentioned in our analysis of the possible working regimes: the system operated with these parameters both conserves the number of excitations and remains within the qubit manifold. This is a bosonic model, where each site can be either empty or occupied by a single particle. The system is then described by the effective Hamiltonian
\begin{equation}\begin{split}
 \hat H_{\rm HCB} = 
 \sum_{i}\half\p{\omega_{\rm q} + \Delta E_{i}}\hat \gs^{z}_{i} - J\sum_{\avg{i,j}}\p{\hat\gs^{+}_{i}\hat \gs^{-}_{j} + \hat\gs^{+}_{j}\hat \gs^{-}_{i}}
\label{eq:HCB}
\end{split}\end{equation}
where $\hat \gs^{z}_{i},\hat \gs^{\pm}_{i} = \hat \gs^{x}_{i} \pm i\hat \gs^{y}_{i}$ are the Pauli $z$ and raising and lowering operators on site $i$.

\begin{figure}[t] 
   \centering
   \includegraphics[scale=1]{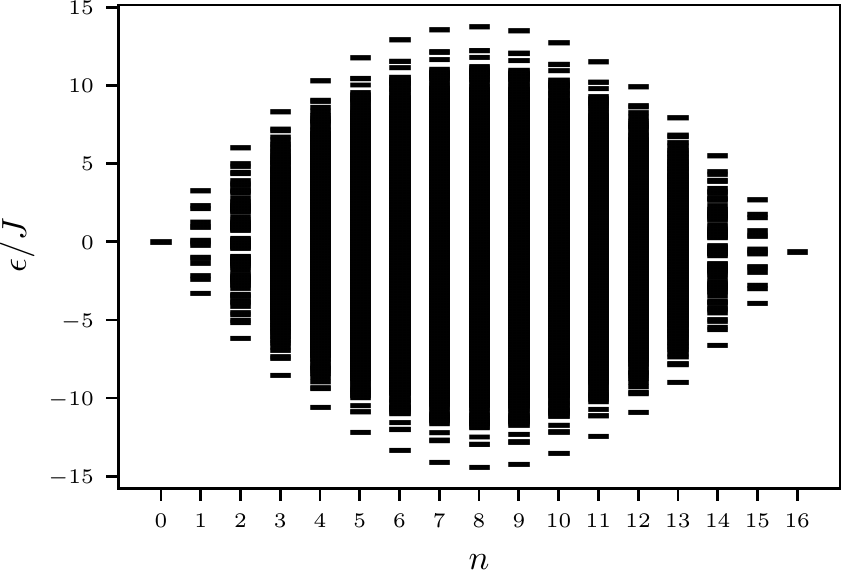}

   \caption{{\bf The spectrum of the the Hard-Core Bose-Hubbard model in the rotating frame.} Here, calculated for a single realization of a $4\times4$ square lattice with nearest-neighbor hopping (see \cref{fig:sketch}) at $\Delta\omega= 0.2J$. 
   The full spectrum comprises 17 distinct sectors with fixed $n$, separated by $\omega_{\rm q}$ with width proportional to $J$. $\gep$ is the rotating frame energy, as defined in \cref{eq:HCBeigenep}.
      }
   \label{fig:spectrot}
\end{figure}

As the Hamiltonian is number preserving, its spectrum decomposes into $N+1$ distinct sectors defined by the total excitation number $n$. Each sector is composed of ${N \choose n}$ levels, defined by their rotating-frame energy $\gep$, with bandwidth proportional to $J$. The eigenstates of \cref{eq:HCB} are then given by $\ket{n,\gep}$  where
\begin{align}
\sum_{i}\half\p{\hat\gs^{z}_{i}+1}&\ket{n,\gep}=n\ket{n,\gep},
\label{eq:HCBeigenn}
\\\p{\hat H_{\rm HCB}-E_{\rm G}}&\ket{n,\gep} = \p{\omega_{\rm q}n + \gep}\ket{n,\gep}, 
\label{eq:HCBeigenep}
\end{align}
where $E_{\rm G}$ is the ground state energy. This spectrum is sketched out in \cref{fig:spectrot}. 

The HCB is difficult to solve except in some specific cases. The 1D chain can be solved through fermionization \cite{Paredes2004,Girardeau1960}, and the case of $n/N \ll 1$ ($n/N \approx 1$) can be understood analytically by perturbative corrections to the free particle (free hole) problem \cite{Schick1971}; both regimes exhibit noninteracting behavior that is much simpler than what we describe below. In addition, small systems can be exactly diagonalized, as we do here for a $4\times4$ lattice. Beyond these limits, research into the model has generally used tensor network methods and focused on the ground state energy \cite{Murg2007,Jordan2009}. An experimental realization of a 2D version of \cref{eq:HCB} can therefore contribute significantly to our understanding of the eigenstates and dynamics of many-body systems, and also the validity and limits of tensor network methods in large systems. Beyond this, as we discuss below, an experimental realization can access the system's entire spectrum.

\begin{figure*}[t] 
   \centering

   \subfloat[\label{fig:rescorr}Correlation length]{\includegraphics[width=0.33\textwidth]{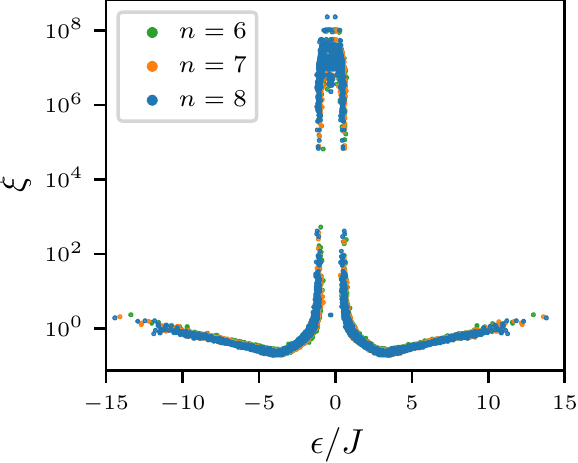}}\hfill
   \subfloat[\label{fig:resent}Entanglement entropy behavior]{\includegraphics[width=0.33\textwidth]{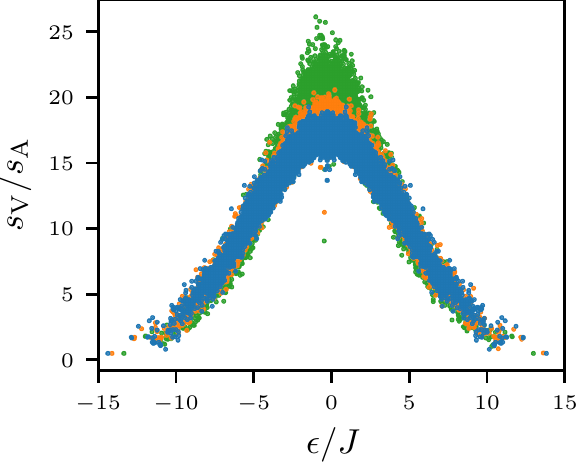}}\hfill
   \subfloat[\label{fig:dsVsA}Sensitivity to disorder]{\includegraphics[width=0.33\textwidth]{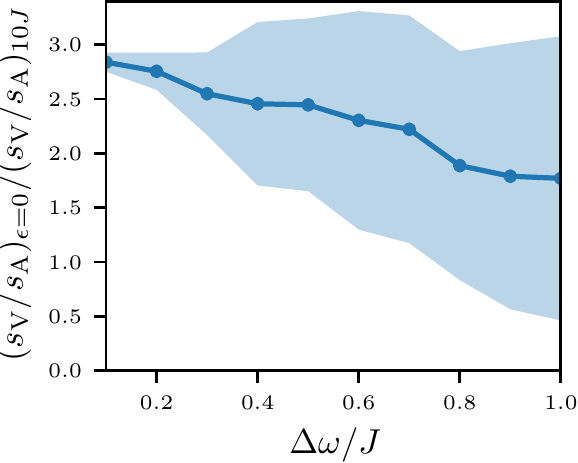}}

   \caption{{\bf Numerical evidence for many-body behavior in the HCB.}
   \protect\subref{fig:rescorr},\protect\subref{fig:resent} Calculated for a single realization of a $4\times4$ square lattice with nearest-neighbor hopping and $\Delta\omega= 0.2J$. 
   We expect the physics of the sectors $n=6,7,8$, near $n\approx N/2$, to be dominated by many-body effects. For each eigenstate in these sectors, we calculate and plot \protect\subref{fig:rescorr} the $C^{\rev{x}}$ correlation length [see \cref{eq:corrlendef}]  and  \protect\subref{fig:resent} the ratio $s_{\rm V}/s_{\rm A}$ between the volume coefficient and area coefficient of the entanglement entropy [see \cref{eq:entent}].
   We observe a clear variation in physics along the spectrum, going from a finite correlation length and area-law behavior of the entanglement entropy at the edges of the band to diverging correlation length and volume-law behavior at the center of the band
   (Note that in a $4\times4$ system the largest separation between qubits is $L=6$, and so $\xi\gtrsim6$ hints at long-distance order).
    We also observe that the behavior is similar for the three sectors with similar $n/N$. The robustness of this signature across filling number enables us to probe entanglement entropy using coherent-like states.
   \protect\subref{fig:dsVsA} For the sector $n=8$, we examine the effects of increased frequency variation, $\Delta \omega$. We plot the ratio between $s_{V}/s_{A}$ at the center of the band ($\gep = 0$) and its edge ($\gep = 10J$), averaged over 10 realization of the disorder. The shaded area gives the range of results over one standard deviation. For $\Delta\omega\le 0.5J$, we can clearly observe the change in physics over the spectrum; for $\Delta \omega > 0.5J$, the variance due to different realizations dominates. 
    }
   \label{fig:HCB}
\end{figure*}

We consider two particular measures of the system's many-body spectrum: the correlation length and the behavior of entanglement entropy for each eigenstate. In \cref{fig:HCB}, we show these quantities exhibit transitions along the spectra within each sector: as we go from the edges of the band to the center, the correlation length grows from finite to infinite, and the entanglement entropy of subsystems evolves from obeying an area-law dependence on the subsystem's size to a volume-law dependence.

\paragraph{Correlation length:} The typical scale beyond which different sites are no longer correlated serves as an order parameter for phases with long-range order~\cite{Altland2010}. The correlation length is a limiting factor for the applicability of tensor-network methods, which can be used only when correlations are finite~\cite{Orus2014}. Having experimental access to the correlation length therefore provides significant insight into the many-body properties of the system.

For our purpose, we define the correlation length in terms of the correlation function
\begin{equation}
C^{\rev{x}}_{i,j}\p{\ket{\psi}} \equiv \bra{\psi}\hat \gs^{\rev{x}}_{i}\hat \gs^{\rev{x}}_{j}\ket{\psi} -\bra{\psi}\hat \gs^{\rev{x}}_{i}\ket{\psi}\bra{\psi}\hat \gs^{\rev{x}}_{j}\ket{\psi}.
\end{equation}
We then extract the correlation length of a state $\ket{n,\gep}$ by fitting $C^{\rev{x}}_{i,j}\p{\ket{\psi}}$ to the form
\begin{equation}
\abs{C^{\rev{x}}_{i,j}\p{\ket{n,\gep}}}^{2}\simeq \mathcal A\p{n,\gep} \exp\br{-\abs{\vec r_{i}-\vec r_{j}}/\xi\p{n,\gep}},
\label{eq:corrlendef}
\end{equation}
over all pairs $i,j$ of nearest neighbors and next-nearest neighbors. Here $\abs{\vec r_{i}-\vec r_{j}} = \abs{x_{i}-x_{j}} + \abs{y_{i} - y_{j}}$ is the Manhattan distance between the sites $i,j$.

We plot the correlation length $\xi$ as a function of eigenstate energy in \cref{fig:rescorr} for a $4\times4$ lattice \rev{near half-filling, where we expect many-body effects to dominate}. As discussed above, we observe it goes from finite and short for states at the edge of the band to effectively infinite for states at its center.

\paragraph{Entanglement entropy:}
For a state with density matrix $\hat \rho$, the entanglement entropy of some subset $X$ of the lattice is the entropy generated when it is severed from the rest of the system,
\begin{equation}
\mathcal S_{X}\p{\hat \rho} = S\p{\hat \rho_{X}\otimes\hat \rho_{\bar X}} - S\p{\hat\rho},
\end{equation}
where $S\p{\hat\gs}$ is the entropy of $\hat \gs$, and ${\hat \rho_{X} = \Tr_{\forall i\notin X}\hat \rho}$, ${\hat \rho_{\bar X} = \Tr_{\forall i\in X}\hat \rho}$ are the reduced density matrices of the subsystem $X$ and the remainder of the lattice, respectively. Note that if the initial density matrix $\hat\rho = \ket{\psi}\bra{\psi}$ is a pure state, $S\p{\hat\rho}$ vanishes while the entropy of both subsystems must be identical, so that
\begin{equation}
\mathcal S_{X}\p{\hat \rho} = 2S\p{\hat \rho_{X}}.
\end{equation}
Throughout this paper, for the purpose of numerical calculations, we use the second R\'enyi entropy,
\begin{equation}
S\p{\hat\rho} = -\log \Tr \hat\rho^{2}.
\end{equation}

The entanglement entropy is a measure of entanglement between different parts of the lattice, and has been an important tool in the study of many-body systems. In particular, there has been significant study of the difference between states where it is proportional to the size of the subsystem $X$ (``volume-law'') and where it is proportional to the size of its boundary (``area-law'')~\cite{Eisert2010}. Volume-law states are also harder to approximate using tensor-network methods.

To describe the growth law for an eigenstate $\ket{n,\gep}$, we extract the parameters $s_{\rm V}$ and $s_{\rm A}$ by fitting the entanglement entropy to the form
\begin{equation}
\mathcal S_{X}\p{\ket{n,\gep}\bra{n,\gep}} \simeq s_{\rm V}\p{n,\gep}V_{X} + s_{\rm A}\p{n,\gep}A_{X},
\label{eq:entent}
\end{equation}
over different lattice subsets $X$. Here $V_{X}$ is the number of sites in $X$ (its ``volume'') and $A_{X}$ is the number of coupling terms between sites in $X$ and the rest of the lattice (its ``area''). The fit parameters can then be understood as
\begin{equation*}\begin{split}
s_{\rm V} & \quad \text{Bulk entanglement entropy per site,}
\\ s_{\rm A} & \quad \text{Boundary entanglement entropy per bond.}
\end{split}\end{equation*}
Thus, the ratio $s_{\rm V}/s_{\rm A}$ determines whether the entanglement entropy obeys an area-law-like or volume-law-like behavior. 

We plot this quantity for a $4\times4$ lattice in \cref{fig:resent}. We see the transition from area-law behavior for states at the edges of the band to volume law behavior at its center. We also see little variation in this behavior between different sectors with similar $n/N$. This allows us to explore the behavior of the entanglement entropy by preparing coherent-like superposition states across multiple sectors, as described below.

\subsubsection*{Measuring entanglement}
Global measures such as the entanglement entropy are key to understanding many-body properties, but observing them in the lab poses experimental challenges. Naively, the entropy of a state is derived from the density matrix $\hat \rho$ and extracting it requires full state tomography. The challenge here is two-fold: first, the number of measurements scales exponentially as $2^{2N}$ \cite{Haah2017}; and second, one must have sufficient control to apply any combination of rotations $\hat\gs^{\pm}_{i}$ to all sites concurrently. 

The situation, however, is not quite so dire. Multiple recent proposals have suggested alternative approaches for measuring non-local observables such as $n$-time correlation functions \cite{Pedernales2014} and the second R\'enyi entropy \cite{vanEnk2012,Elben2019,Elben2018,Vermersch2019}. These proposals substitute random unitaries for the full set of rotations mentioned above, easing the control requirements. They also require fewer unitaries than does full state tomography, though the number of measurements needed still scales exponentially with system size. We note, though, that even as the total size of the system increases, the scaling coefficients $s_{\rm V},s_{\rm A}$ can be determined from the entanglement entropy of fixed-size subsystems (e.g.~a block of sites of size  $3\times3$ and all its subsystems), leaving the required number of measurements constant even if we increase $N$.

\subsubsection*{Frequency variance}
As noted above, the emergence of many-body behavior requires relatively uniform qubit frequency, $\Delta\omega \lesssim J$. In \cref{fig:dsVsA}, we quantify the tolerable amount of variation for the metrics discussed here. We do so by calculating the behavior of the entanglement entropy at the center of the band and at its edge at varying disorder strength, averaged over multiple realizations of the lattice. We find that up to $\Delta \omega \approx 0.5J$, one can observe distinctly different physics in different parts of the spectrum. At larger frequency disorder, the variation between lattice realizations dominates this effect.

\subsection*{Proposal for transmon implementation}

The transmon qubit~\cite{Koch2007} is a natural building block for the implementation of the HCB with a superconducting circuit. It behaves as a weakly non-linear oscillator with a fundamental transition frequency in the range of $\omega_{\rm q}/2\pi\sim\SI{5}{GHz}$. Each lattice site is represented by a single transmon qubit, with the local site energy corresponding to the qubit transition frequency $\omega_{\rm q}$. 

The anharmonicity of the transmon qubit is negative, typically in the range of ${A/2\pi\sim\SI{-250}{MHz}}$ or $\sim5\%$ of its frequency~\cite{Koch2007}. The self-Kerr non-linearity of the transmon Hamiltonian maps directly onto the on-site interaction term in the Bose-Hubbard model \cite{Ma2019}.
Since the hard-core Bose-Hubbard model operates in a regime where $J/\abs{A}\ll 1$ (Mott insulator phase), the population of the same lattice site with two or more particles is strongly suppressed due to the presence of the self-Kerr term, irrespective of its sign.
One may note that for large enough lattices, the kinetic energy may reach the scale of the anharmonicity $\gep \propto NJ \sim \abs{A}$. Generally, this effect can be treated as a perturbative correction to the hard-core approximation, as we expect to see only a small number of sites out of a large occupation $\propto N$ in the forbidden state.

It is straightforward to connect transmon qubits via capacitive coupling \cite{Ma2019}, leading to the hopping term in the Bose-Hubbard model with nearest-neighbor coupling energy $J$. Typical achievable coupling strengths are tens of megahertz, rendering the qubit-qubit interaction well within the strong coupling regime $J>\Gamma$, where $\Gamma$ denotes the qubit decoherence rate. 
Contemporary transmon qubits feature reproducible coherence times in the range of $\SIrange{20}{100}{\micro s}$~\cite{Kjaergaard2019}, corresponding to ${\Gamma/2\pi \lesssim \SI{10}{kHz}}$.

As discussed above, an experimental implementation operating in the regime $J/\abs{A}\ll 1$ suppresses transitions to the second and higher levels, and implements the HCB. In order to observe many-body physics, we generally require the qubit lifetime to be much longer than the characteristic time scale for information to traverse the system, $1/\Gamma \gg L/J$ where $L$ is the number of hops to go across the system (its length). With five orders-of-magnitude in separation, $\abs{A} \gtrsim 10^{5}\Gamma$, this is easily achievable with transmon lattices of 100 qubits or more. Generally the sweet spot in this case is ${\p{J\sim \sqrt{\abs{A}\times \Gamma L}}/2\pi \sim \SI{1}{MHz}}$.

\rev{
Qubit coherence in the proposed transmon HCB lattice is expected to be at the level of individual state-of-the-art transmon qubits \cite{Kjaergaard2019}, limited by a combination of material defects \cite{Oliver2013} and parasitic coupling to stray modes in the sample package \cite{Lienhard2019}. Scaling to a larger number of qubits typically requires a chip and sample package of larger dimensions, with the risk of introducing additional parasitic modes at frequencies at or close to the qubit frequencies and therefore impairing qubit performance. In previous implementations of arrays with $24$ and $53$ qubits, energy relaxation times averaged at around $\SI{15}{\micro s}$ \cite{Arute2019} and $\SI{10}{\micro s}$ \cite{Ye2019}.
}
\subsubsection*{\rev{Frequency control}}

Fabrication variations translate to variations in transmon transitions which may exceed ${\SI{200}{MHz}}$ \cite{Gambetta2017}, yielding disorder in the emulated model on the order of $\Delta\omega\sim\abs{A}$. To compensate for such variation, we consider a lattice of frequency-tunable transmon qubits. This is achieved by replacing the single Josephson junction of the qubit with a dc-SQUID, facilitating a frequency tunability of several $\SI{}{GHz}$. In an experiment, this enables one to tune the individual qubit frequencies mutually on resonance (to within their spectral linewidth \cite{Ma2019}).

Individual frequency control requires $N$ slow (dc) control lines for flux biasing each of $N$ qubits. Such low-frequency wiring can be straightforwardly routed in dilution refrigerators and connected to the sample package in large numbers, as the necessary connectors are compact and bulky attenuation at multiple temperature stages is not required.

Frequency variation in the lattice is mitigated experimentally by calibrating the (dc) flux cross-talk matrix, containing information about the frequency shift of qubit $i$ responding to a flux bias applied to bias line $j$ (${1\le i,j\le N}$). In large lattice implementations, qubits are physically located far away from flux bias lines of other qubits. By taking into account only nearest neighbor and next-nearest neighbor parasitic flux coupling, the resulting flux cross-talk matrix is sparse, reducing the number of matrix elements from $O\p{N^2}$ to $O\p{N}$.

In general, flux cross-talk calibration requires the measurement of sections of all $N$ qubit spectra while consecutively biasing each of the $N$ flux control lines. As the spectra can be measured simultaneously with multiplexed readout, this requires $O\p{N}$ individual measurement scans and therefore scales linearly with lattice size.

Additionally, dynamic (ac) flux control allows for rapid frequency tuning of the qubits. By detuning a qubit away from its neighbors, we can effectively decouple it from the lattice. For example, in a square lattice, system dynamics can be entirely frozen out, enabling state preparation and readout, by detuning every other qubit in a checkerboard pattern, where all ``white'' qubits remain at the original frequency and all ``black'' qubits are shifted.

This scenario requires $N/2$ qubits to be equipped with fast flux lines, such that even a large lattice of size $10\times10$ requires only $50$ flux control lines. Assuming individual bias lines used, enabling full control on each qubit, the number of required coaxial lines is still moderate compared with recent implementations using $50$ and $200$ coaxial control lines for a $24$-qubit and $50$-qubit chip, respectively \cite{Arute2019,Ye2019}.

\begin{figure*}[t] 
\center

   \parbox[t]{0.35\textwidth}{
   	\subfloat[\label{fig:implschem}Schematic for a floating-transmon implementation of a 2D HCB]{\includegraphics[width=0.35\textwidth]{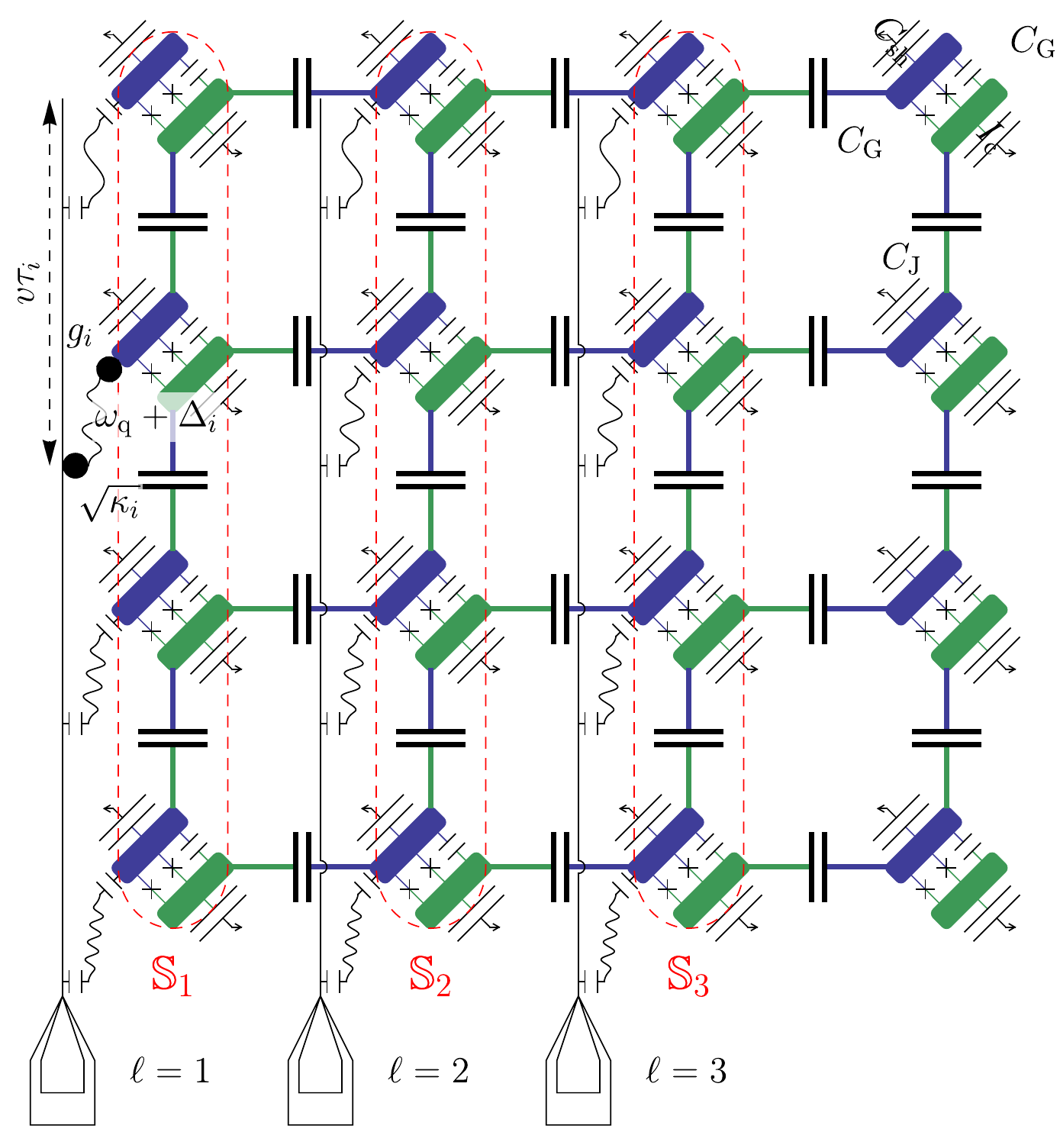}}
   }
   \hfill
   \parbox[t]{0.21\textwidth}{\centering
   \vspace{15pt}
   
   \subfloat[\label{fig:circfl}Floating transmon]{\includegraphics[width=0.2\textwidth]{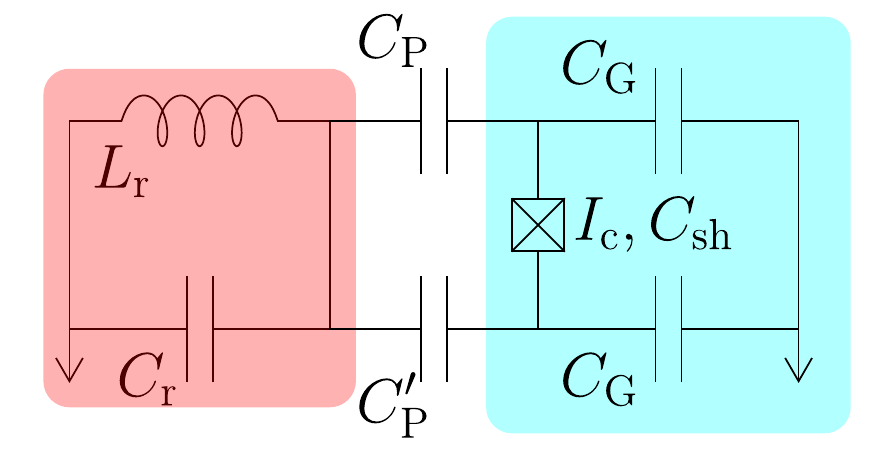}}

   \subfloat[\label{fig:circgr}Grounded transmon]{\includegraphics[width=0.2\textwidth]{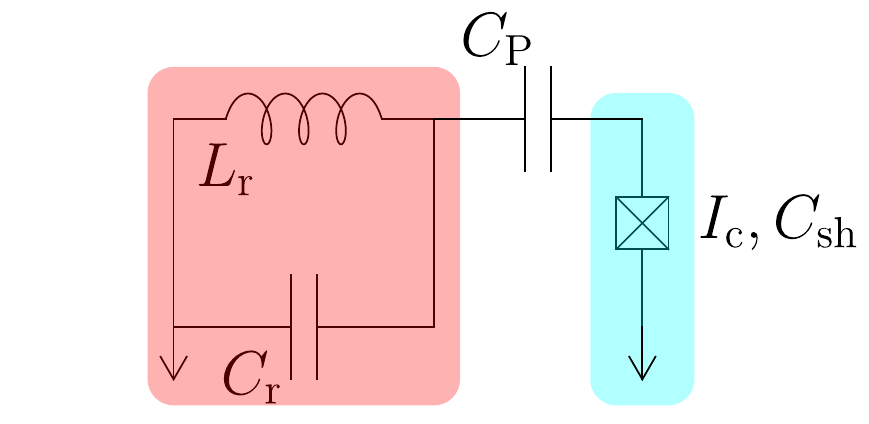}}
   }
   \hfill
   \parbox[t]{0.4\textwidth}{
   	\subfloat[\label{fig:Ceff}Parasitic coupling of a resonator mode to a floating and a grounded transmon]
	{\includegraphics[width=0.4\textwidth]{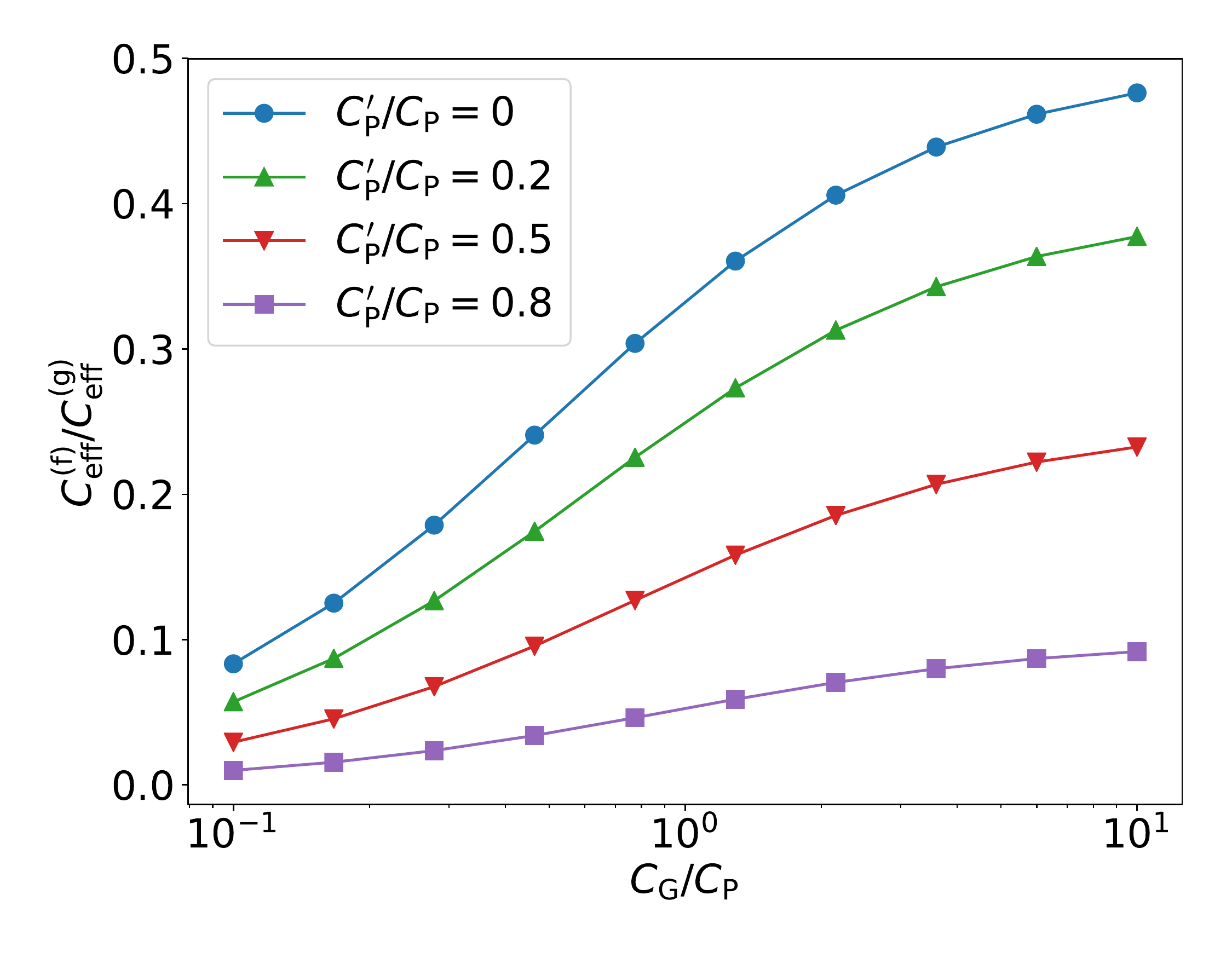}}
   }

\caption{\rev{\bf Implementation of the 2D HCB based on floating transmon qubits.}
\protect\subref{fig:implschem} Schematic circuit implementation of the 2D Bose-Hubbard grid using transmon qubits with floating electrodes. 
	The relevant capacitances are the direct shunting capacitance $C_{\mathrm{sh}}$ between the qubit electrode pads, the capacitance $C_{\rm G}$ of each pad to the ground, and the coupling capacitance $C_{\mathrm{J}}$ between electrodes of adjacent qubits. 
	By maintaining this coupling between left electrodes (blue) and right electrodes (green), the phase of the mutual couplings around a four-qubit plaquette sums to zero. A different choice could be made to create an effective gauge field with flux $\pi$.
	The two Josephson junctions form a dc-SQUID with total effective critical current $I_{\rm c}$, which allows the qubit frequency to be tuned.
	We also show the dispersive readout and control scheme. Here, three control lines lines ($\ell=1,2,3$) are coupled to groups of qubits (marked $\mathbb S_{1},\mathbb S_{2},\mathbb S_{3}$) via detuned resonators. The different resonators on each line have different frequencies, and thus detunings from the uniform qubit frequency. Each resonator is characterized by its frequency $\omega_{\rm q}+\Delta_{i}$, position along the input line $\tau_{i}\times$speed of light, linewidth $\gk_{i}$, and qubit coupling strength $g_{i}$.
	\protect\subref{fig:circfl}, \protect\subref{fig:circgr} Schematic circuit representation for a resonator (shaded red) parasitically coupled to a transmon qubit (shaded blue) with \protect\subref{fig:circfl} floating electrodes and \protect\subref{fig:circgr} one grounded electrode. 
	\protect\subref{fig:Ceff} The effective coupling capacitance $C_{\mathrm{eff}}^{(\mathrm f)}$ for floating transmon qubits with various combinations of parasitic couplings $C_{\rm P}$ and $C_{\rm P}\pr$ (colors). $C_{\mathrm{eff}}^{(\mathrm f)}$ is suppressed compared with a grounded transmon with the same parasitic capacitance.
	}
\label{fig:impl}
\end{figure*}

\subsubsection*{Implementation with floating transmon qubits}

\Cref{fig:implschem} shows a possible circuit implementation of the 2D HCB based on a grid of transmon qubits each consisting of two floating electrodes \cite{Chang2013,Corcoles2015,Braumueller2016,Reagor2018}, in contrast to recent realizations where one of the electrodes is grounded (e.g.~Xmon qubits \cite{Barends2013}).
In circuit designs with an increasing number of qubits, circuit elements can be proximal or even overlap when using cross-over fabrication techniques~\cite{Rosenberg2017}. This may result in unwanted spurious coupling, referred to as cross-talk. Such spurious couplings can exist between signal lines, readout resonators, and qubits. In order to minimize this effect, it is advantageous to confine electric fields by decreasing the mode volume; this however comes at the expense of an increased electric field strength, leading to enhanced surface defect loss~\cite{Martinis2005,Oliver2013}. The floating layout is advantageous since it suppresses parasitic couplings in the circuit. 

To see this, we compare the parasitic coupling of a resonator mode to a floating and a grounded transmon.
We assume a (parasitic) capacitive coupling $C_{\rm P}, C_{\rm P}\pr$ between the resonator and the electrodes of floating transmon (circuit diagram in \cref{fig:circfl}), or $C_{\rm P}$ to the electrode of a grounded transmon (\cref{fig:circgr}). 
While the coupling capacitance for the grounded transmon is simply 
\begin{equation}
C_{\rm eff}^{\rm \p{g}}=C_{\rm P},
\end{equation}
the effective coupling capacitance between resonator and the floating transmon depends on the parasitic capacitances $C_{\rm P},C_{\rm P}\pr$ as well as the capacitance to the ground $C_{\rm G}$. Assuming without loss of generality $C_{\rm P}\pr \le C_{\rm P}$, circuit analysis (see the Methods section)
yields an effective coupling capacitance
\begin{equation}
C_{\rm eff}^{\rm \p{f}}=\frac{C_{\rm G}\p{C_{\rm P}-C_{\rm P}\pr}}{2C_{\rm G}+C_{\rm P}+C_{\rm P}\pr} \le \frac{C_{\rm P}}{2}=\frac{C_{\rm eff}^{\rm \p{g}}}{2}
\label{eq:Cefff}
\end{equation}
We note $C_{\rm eff}^{\rm \p{f}}\le \half C_{\rm eff}^{\rm \p{g}}$ generically and $C_{\rm eff}^{\rm \p{f}}\ll C_{\rm eff}^{\rm \p{g}}$ if $C_{\rm P}\pr\approx C_{\rm P}$. This corresponds to an effective confinement of electric fields, which is advantageous in larger and more complex circuits. The relation of the effective coupling capacitances $C_{\rm eff}^{\rm \p{f}}/C_{\rm eff}^{\rm \p{g}}$ is plotted in \cref{fig:Ceff} for typical parameters $C_{\rm P},C_{\rm P}\pr,C_{\rm G}$. The argument remains valid if the capacitances of the two transmon electrodes to ground are not identical.

Another potential benefit of the floating transmon design is that it provides a tuning knob for the strength of long-range interactions within the lattice. In particular, the coupling range between non-adjacent qubits can be adjusted by controlling $C_{\rm G}/C_{\mathrm{sh}}$\rev{\cite{unpub}}, the ratio of the qubit shunt capacitance to the capacitance of the two pads to the ground. For the implementation of the HCB Hamiltonian, \cref{eq:HCB}, next-nearest neighbor couplings must be suppressed, which can be achieved in the limit where $C_{\rm G}\gg C_{\mathrm{sh}}$, but the use of floating transmons opens the possibility of exploring models with non-local interactions in the future.

Another strategy to mitigate unwanted cross-talk is to physically separate circuit elements by introducing a multi-layer chip layout (3D integration) \cite{Rosenberg2017}. This approach is particularly beneficial in the implementation of a 2D grid of qubits, since the circuit topology prevents in-plane access to interior qubits. In a planar layout, this can be resolved by using airbridges to cross over signal lines \cite{Chen2014}, but these are naturally prone to unwanted cross-talk. The 3D integration approach allows for a separation of coherent elements (qubits) on one layer and signal lines on another layer, with their respective electric fields well separated. Couplings between qubit and control or readout lines are achieved via a flip-chip approach and connectivity to the other substrate surface is facilitated by through-silicon vias (TSV), which are low-loss superconducting trenches etched inside the silicon substrate~\cite{Rosenberg2017,Yost2019}.

\subsubsection*{Qubit readout}

Individual qubit readout and control in devices with only few qubits \rev{can be} achieved by connecting a separate signal line to each qubit. For a QMBS-style device with a large number of qubits, this approach is limited by the available number of signal lines as well as by geometric constraints. 
Instead, efficient multiplexed readout can be performed by coupling multiple qubits to a single signal line through individual dispersive readout resonators with frequencies spaced at intervals large compared to their linewidths \cite{Blais2004,Heinsoo2018}. We sketch out an example of this setup in \cref{fig:implschem}.

As implied by the color coding in \cref{fig:implschem}, signal lines must cross qubit pads or qubit coupling elements in a planar circuit implementation in order to reach qubits inside the lattice, leading to experimental challenges. A possible strategy to address this issue is the use of 3D integration techniques~\cite{Rosenberg2017}.

The particulars of dispersive readout for individual qubits are well established. 
The challenge in reading out a large, degenerate array of qubits is the interplay between measurement and the ongoing dynamics. To get a snapshot of the system at a particular time, we must generally measure the qubits on a time scale $T_{\rm meas} \ll 1/J$, or else freeze the dynamics.

For a homodyne measurement, typical measurement time scales as~\cite{Gambetta2008}
\begin{equation}
T_{\rm meas} \gtrsim \frac{1}{\gk} + \frac{\gk^{2} + \p{\chi/2}^{2}}{\gk \bar n \chi^{2}},
\label{eq:Tm}
\end{equation}
where $\gk$ is the resonator linewidth, $\chi$ its dispersive shift between qubit states $g,e$ and $\bar n$ is the mean number of photons in the cavity during readout. There are two limiting factors this readout speed: cavity occupation must remain below the critical photon number in order to ensure to operate in the linear dispersive regime, and the induced Purcell decay of the qubit, $\gamma_{P}$, must remain small \cite{Houck2008,Walter2017},
\begin{equation}
\bar n \ll \bar n_{\rm crit} = \frac{A}{4\chi}, \qquad \gamma_{P} = \eta_{\rm PF}\frac{\gk \chi}{A} \ll J/L.
\end{equation}
Here $\eta_{PF}$ accounts for protection resulting from a Purcell filter \cite{Walter2017}, and $L$ is the maximum distance between any two qubits. We have taken the anharmonicity $\abs{A}$ to be much smaller than the qubit-resonator detuning.
If we keep fixed
\begin{equation}
\bar n/\bar n_{\rm crit} = \gve_{1}, \qquad L\gamma_{\rm P}/J = \gve_{2},
\end{equation}
then measurement time is 
\begin{equation}
JT_{\rm meas} \gtrsim \frac{J}{\gk} + \frac{4\eta_{\rm PF}L}{\gve_{1}\gve_{2}}\frac{\gk^{2}}{JA^{2}}.
\label{eq:Tmopt}
\end{equation}

We find that the operating regime for measurement depends on the ratio $J/\abs{A}$, favoring a smaller $J$. The design is limited, however by the factors we have previously discussed. The hopping energy must exceed the frequency disorder across different qubits, $J\gtrsim\Delta\omega$; with individual (dc) flux bias lines, one can reasonably expect to achieve $\Delta\omega/2\pi \approx  \SI{100}{kHz}$, independent of lattice size. The rate must also be fast enough to allow information to travel across the entire system before decoherence kicks in; at a conservative qubit lifetime of $T_{1} \approx \SI{10}{\micro s}$ this translates into the requirement ${J/2\pi \gg L\times \SI{15}{kHz}}$. Thus, a larger $10\times10$ lattice, with $L=20$, requires $J/2\pi \gtrsim \SI{3}{MHz}$, $J/\abs{A}\gtrsim 0.01$. 

We find that one of several experimental approaches can be taken:
\begin{itemize}
\item If $J  \lesssim 0.03\sqrt{\gve_{1}\gve_{2}/\eta_{\rm PF} L}\abs{A}$, we can choose system parameters, and in particular $\gk>J$, such that $ JT_{\rm meas}  \lesssim  \pi/10$. In this case, we can easily read out the state of the system faster than it evolves. This regime can be reached\footnote{Taking a typical $\gve_{1}=\gve_{2}=0.2$} by using narrow Purcell filters, $\eta_{\rm PF} \le 0.01$, and large bandwidth cavities, $\gk/2\pi \gtrsim \SI{20}{MHz}$. Multiplexed non-demolition qubit readout  with similar parameters was demonstrated in less than $T_{\rm meas} = \SI{50}{ns}$ \cite{Walter2017}.

The experimental overhead for this approach is large in bigger lattices. As the Purcell filters must be spectrally very narrow, only one cavity can be brought in direct resonance with each filter, and so each qubit needs a readout resonator and a separate Purcell filter. Cavity frequencies must be spaced sufficiently far apart, at intervals of $\Delta \gtrsim \SI{100}{MHz}$. Another downside to this approach is that narrow filters, while increasing the qubit lifetime, make it hard to drive the qubits through the readout line. As we discuss below, we find that this is a useful tool in preparing states that explore the system's many-body properties, and if the readout cannot be used in this way separate drive lines would be necessary.

\item If $0.03\sqrt{\gve_{1}\gve_{2}/L}\abs{A} \lesssim J\ll \gve_{1}\abs{A}$, measurement speed is limited. However, if the Stark shift generated by driving the cavity detunes the measured qubit away from the lattice, $\gd\omega_{\rm q} \sim \bar n \chi\sim \gve_{1}\abs{A}\gg J$, its state is frozen and we can once again read out a snapshot at a given time.

A design of this form would call for $\p{\chi \approx -\gk}/2\pi \approx \SI{5}{MHz}$. If neighboring qubits are to be measured simultaneously, the driving pulses must be carefully calibrated to maintain a frequency detuning, which means the protocol may not be robust.

\item Finally, if $J\gtrsim \gve_{1}\abs{A}$, we necessarily have $T_{\rm meas}\gtrsim 1/J$.
This is the weak continuous measurement regime \cite{Clerk2010}, and the amount of information that can be extracted about the system is reduced: we would not, for example, be able to obtain the probability statistics required to measure the entropy of a state. In this regime, full readout can be enabled by turning off interactions, either directly \cite{Yan2018a}, or by making use of frequency-tunable qubits. 

We can effectively freeze out the interactions between the qubits by mutually detuning their frequencies, essentially shifting the system into the individual particle regime. 
Note that we do not need an infinite array of frequencies, as only coupled qubits must be detuned from each other. In a square lattice, qubits can be detuned in a checkerboard pattern, as described above. This freezeout can be achieved by attaching fast flux lines to $N/2$ qubits, requiring comparable or reduced overhead to the use of individual Purcell filters, similar to previously realized setups \cite{Arute2019,Ye2019}. This method also allows for more flexibility in measuring observables other than $\hat \gs^{z}$, as rotation pulses can be applied to the qubit between detuning and measurement, possibly through the cavity array, as described below.

\end{itemize}

\subsubsection*{Microwave control} 
While the resonator configuration discussed above enables selective readout of specific or all qubits, it does not facilitate individual qubit control with microwave drives when all the qubits in the lattice are degenerate.

This issue can be overcome in several ways. Most directly, it may be useful to couple control lines to a single or few specific qubits to allow for direct microwave control, e.g., to prepare a certain initial state in the lattice. Alternately, the use of tunable qubits -- which we have suggested above for the purpose of a freeze-out prior to qubit readout -- allows one to address an individual qubit or a subset of qubits if they are detuned in frequency away from the otherwise degenerate lattice.

In addition, the readout layout described above can be used to effect a specific form of system-wide driving. We note, when a signal line $\ell$ is driven at near resonance, at ${\omega_{\rm d}\approx \omega_{\rm q}}$, the effective Hamiltonian becomes
\begin{equation}\begin{gathered}
\hat H \to \hat H_{\rm HCB} + \tilde g\p{e^{-i\omega_{\rm d}t}\hat \Sigma_{\ell}\dg  + e^{i\omega_{\rm d}t}\hat \Sigma_{\ell}},
\label{eq:Hdr}
\end{gathered}\end{equation}
where the driving operator is given by
\begin{equation}\begin{gathered}
\hat \Sigma_{\ell} = \sum_{i\in \subS_{\ell}}\tilde \ga_{i}\hat \gs^{-}_{i}.
\label{eq:DriveOp}
\end{gathered}\end{equation}
Here, the summation is over the set of qubits $\mathbb S_{\ell}$ coupled to the signal line $\ell$ (see \cref{fig:implschem}), $\tilde \ga_{i}$ is the effective relative coupling to that qubit, determined by the resonator's parameters, and $\tilde g$ is a coupling energy proportional to the driving strength. See the Methods section
for the derivation of this operator and the values of $\tilde g, \tilde \ga_{i}$.

While the set of operators $\hat\Sigma_{\ell}$ does not allow us full control of the system, driving at different strengths or for different lengths of time allows us access to a set of defined unitary transformations. \rev{As mentioned above, }this would allow the measurement of quantities such as the entropy of a subsystem \cite{vanEnk2012,Elben2019}. As we discuss below, it also enables the preparation of many-body states whose nature is determined by the detuning of the drive from the qubit frequency and can be used to probe the spectrum of the system.

\subsection*{Coherent-like states}

As we have seen, the most interesting behavior of the HCB is manifest in the finite-excitation density sectors where ${0<n/N<1}$. Within these sectors, energy eigenmodes vary in their behavior between the edges of the band and its center, exhibiting many-body properties such as different entanglement entropy laws. To study these properties, we must be able to prepare such states, which is challenging. 
In our proposed implementation, state preparation can be performed by applying drive pulses that reach the qubits via the readout resonators. To prepare a specific eigenstate, we would not only have to tailor a series of specific pulses, but also know the wavefunction of the prepared state, negating the premise of a quantum simulator to access states which are not understood theoretically.

\begin{figure*}[tb] 
   \centering
   \hfill\subfloat[\label{fig:prepoft}Evolution of the state under weak driving]{\includegraphics[width=0.45\textwidth]{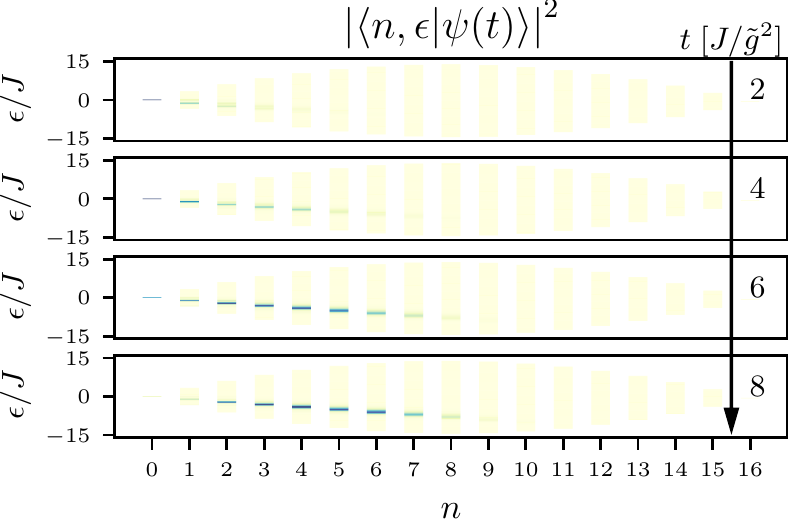}}\hfill
   \subfloat[\label{fig:prepofg}Prepared state at varying driving strength]{\includegraphics[width=0.45\textwidth]{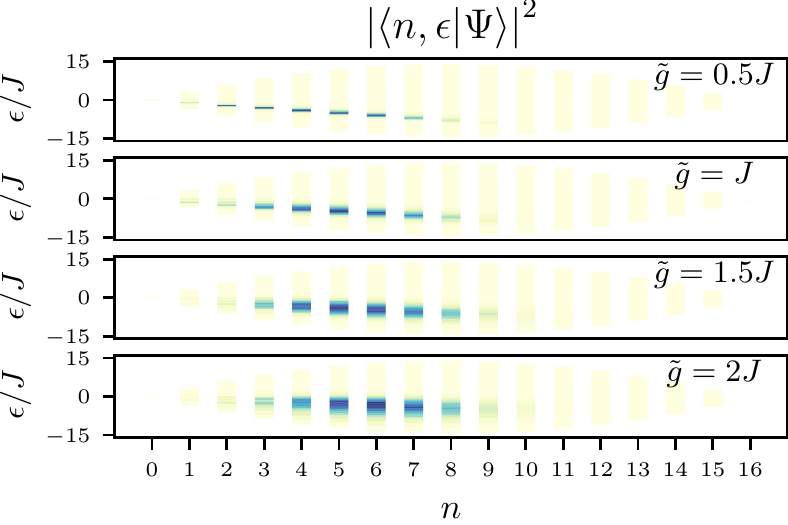}}\hfill\hfill
   \caption{{\bf Coherent-like state preparation.}
   We show here numerical results for the same $4\times4$ system shown in \cref{fig:HCB},  $\Delta\omega = 0.2J$, with a driving term described by ${\hat H_{\rm dr} = \tilde g\p{e^{i\p{\omega_{\rm q}+\gd}t}\hat \Sigma + e^{-i\p{\omega_{\rm q}+\gd}t}\hat \Sigma\dg}}$, where $\hat \Sigma$ is as described in \cref{eq:DriveOpAppx,eq:driveopdetailedg,eq:driveopdetailedga,eq:driveopdetailedK}, taking realistic experimental parameters.
   We plot the overlap of the state with different eigenvalues, $\abs{\braket{n,\gep}{\psi\p{t}}}^{2}$ at different times, and with different driving strength (arbitrary scale, darker colors denote greater overlap). Here, we drive the system at $\gd = -J$.
   \protect\subref{fig:prepoft} we plot the evolution of the state from the initial $\ket{\psi\p{0}} = \ket{0,0}$ for very weak driving, $\tilde g=0.5J$. We see that at any time the state can be described by a superposition of eigenstates $\ket{n,\gd \times n}$, as discussed around \cref{eq:psioft}.
   \protect\subref{fig:prepofg} we plot the prepared state $\ket{\Psi} = \lvert\psi(t = 8\times J/\tilde g^{2})\left.\right>$ at varying $\tilde g$. For stronger driving, the energy width of the prepared state grows as $\Delta E\propto \tilde g$.
    }
   \label{fig:stateprep}
\end{figure*}

Here, we propose an alternate route to observing the spectral properties of the HCB.
Instead of preparing a specific known eigenstate, we apply a weak drive using the operators of \cref{eq:DriveOp} at some detuning from the joint qubit frequency. This prepares the lattice in a coherent-state-like superposition of eigenstates in multiple $n$ sectors, but with definite kinetic energy within each sector. This strategy of extracting many-body properties is robust with regards to experimental control limitations on chip.

\subsubsection*{Preparing coherent-like states}

To understand this process, we begin by rewriting the Hamiltonian of \cref{eq:HCB} in its eigenmode basis,
\begin{equation}
\hat H_{\rm HCB} = \sum_{n=0}^{N}\intrm{d\gep}\rho^{n}_{\gep}\;\p{\omega_{\rm q}n + \gep}\ket{n,\gep}\bra{n,\gep}
\end{equation}
where $\ket{n,\gep}$ are the eigenstates of \cref{eq:HCBeigenn} and $\rho^{n}_{\gep}$ is the density of states for the sector with $n$ excitations. Then, we rewrite the driving operator of \cref{eq:DriveOp} in the same basis,
\begin{equation}\begin{split}
\hat \Sigma_{\ell}\dg =  & \sum_{n} \intrm{d\gep d\gep\pr}  
  \rho^{n+1}_{\gep}\rho^{n}_{\gep\pr}\bra{n+1,\gep\pr} \Sigma_{\ell}\dg\ket{n,\gep} \ket{n+1,\gep}\bra{n,\gep\pr}.
\end{split}\end{equation}

Consider first the perturbative limit, where the driving is very weak compared with the energy spacing,
\begin{equation}
\forall n,\gep,\gep\pr : \quad\abs{\tilde g \bra{n+1,\gep\pr} \Sigma_{\ell}\dg\ket{n,\gep}}^{2} \rho^{n+1}_{\gep}\rho^{n}_{\gep\pr} \ll 1.
\label{eq:appxdrlldE}
\end{equation}
In this case, the driving operator will couple only eigenstates differing exactly by the detuning, 
\begin{equation}
\gep\pr - \gep = \gd = \omega_{\rm d} - \omega_{\rm q},
\end{equation}
and we can approximate it as a combination of defined-energy raising operators
\begin{equation}\label{eq:appxdrop}
\begin{gathered}
e^{-i\omega_{\rm d}t}\hat \Sigma_{\ell}\dg \approx \intrm{d\gep} e^{-i\hat H_{\rm HCB} t}\hat A_{\gep}\dg e^{i\hat H_{\rm HCB} t},
\\ \hat A_{\gep}\dg = \sum_{n} \sqrt{\rho^{n+1}_{\gep_{n+1}}\rho^{n}_{\gep_{n}}}\bra{n+1,\gep_{n+1}} \Sigma_{\ell}\dg\ket{n,\gep_{n}}\ket{n+1,\gep_{n+1}}\bra{n,\gep_{n}},
\end{gathered}\end{equation}
where $\gep_{n} = \gep + n\times \gd$. Observe that each $\hat A_{\gve}$ couples a subset of eigenstates of the form $\ket{n, \gep + n\times \gd}$. In the spectrum outlined in \cref{fig:spectrot}, these can be identified as the states sitting on a line with slope $\gd$ and intersecting $n = 0$ at $\gep$.

\begin{figure*}[t] 
   \centering
   \subfloat[\label{fig:prepofd}Prepared wavefunction]{\includegraphics[width=0.3\textwidth]{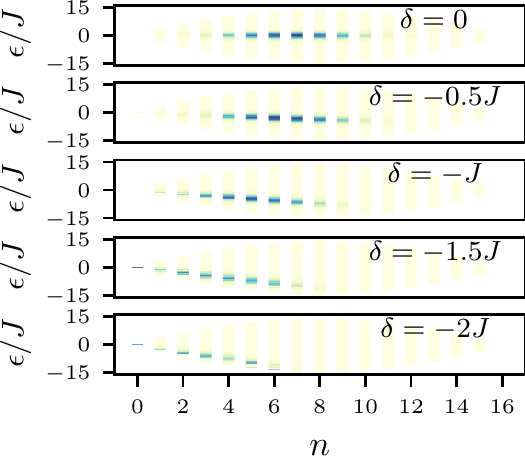}}\hfill
   \subfloat[\label{fig:prepmeasxi}Correlation length]{\includegraphics[width=0.33\textwidth]{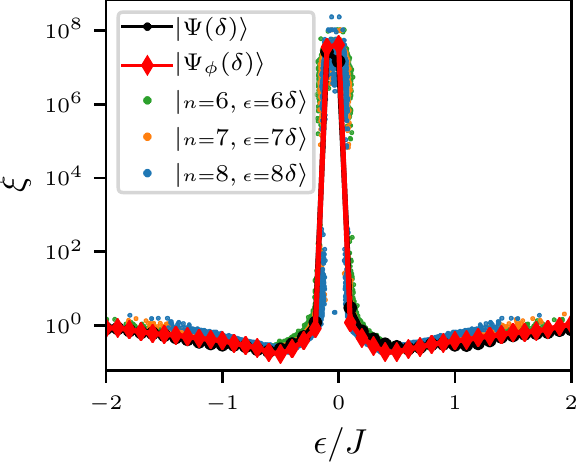}}\hfill
   \subfloat[\label{fig:prepmeasent}Entanglement entropy behavior]{\includegraphics[width=0.33\textwidth]{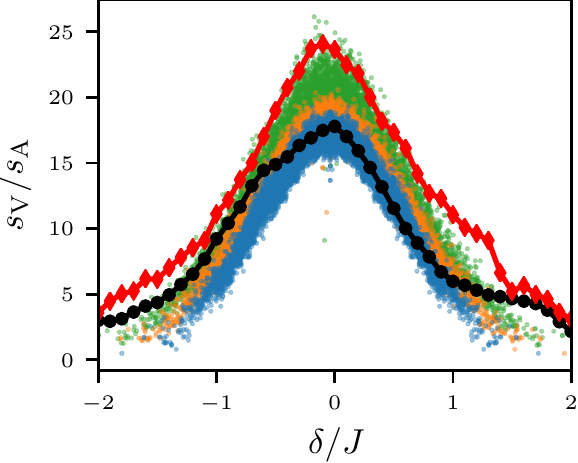}}
   
   \caption{{\bf Probing many-body properties of the HCB with coherent-like states.} We show here numerical results for the same $4\times4$ system shown in \cref{fig:HCB}, $\Delta\omega= 0.2J$, with the driving ${\hat H_{\rm dr} = \tilde g\p{e^{i\p{\omega_{\rm q}+\gd}t}\hat \Sigma + e^{-i\p{\omega_{\rm q}+\gd}t}\hat \Sigma\dg}}$, as in \cref{fig:stateprep}, applied for time $t = 8\times J/\tilde g^{2}$ to prepare the state. Here, we maintain the driving strength $\tilde g=J$ and vary over the detuning $\gd$.
   \rev{$\ket{\Psi}$ is prepared with $\hat \Sigma$ is as described in \crefrange{eq:DriveOpAppx}{eq:driveopdetailedK}, as in \cref{fig:stateprep}, while $\ket{\Psi_{\phi}}$ is prepared with $\hat \Sigma = \sum e^{i\phi_{i}}\hat \gs^{-}_{i}$ for uniformly distributed, random $\phi_{i}$.}
   \protect\subref{fig:prepofd} We plot the overlap of the prepared state with different eigenvalues, $\abs{\braket{n,\gep}{\Psi_{1}}}^{2}$ (arbitrary scale, darker colors denote greater overlap). We see that different values of $\gd$ access different parts of the many-body spectrum.
   \protect\subref{fig:prepmeasxi},\protect\subref{fig:prepmeasent} We compare the many-body properties of the two prepared wavefunctions at various values of the detuning ($\ket{\Psi\p{\gd}}$, black and red lines) to those of the equivalent eigenmodes we expect it to be composed of (colorful dots). These are reproduced from \cref{fig:HCB} with the energy axis rescaled for comparison. We find remarkable agreement \rev{both for 
   \protect\subref{fig:prepmeasxi} the correlation length [see \cref{eq:corrlendef}] and }
   \protect\subref{fig:prepmeasent} the ratio $s_{\rm V}/s_{\rm A}$ between the volume coefficient and area coefficient of the entanglement entropy [see \cref{eq:entent}] for the prepared states.
    }
   \label{fig:stateprepmeas}
\end{figure*}

Thus, if we initialize the system in the ground state,
\begin{equation}
\ket{\psi\p{t=0}} = \ket{0,0},
\end{equation}
it is affected only by $\hat A_{0}, \hat A_{0}\dg$. Inserting the operators of \cref{eq:appxdrop} into the Hamiltonian of \cref{eq:Hdr}, we find at later times it has a form reminiscent of a coherent state,
\begin{equation}\begin{split}
\ket{\psi\p{t}} & \approx e^{-i\hat H_{\rm HCB}t}\exp\br{-i\tilde g\p{ \hat A_{0} + \hat A_{0}\dg}t}\ket{0,0}.
\end{split}\end{equation}
While this wavefunction is difficult to evaluate theoretically, it is composed only of states of a defined energy, $\ket{n,n\times \gd}$, i.e.
\begin{equation}\begin{split}
\ket{\psi\p{t}} & \approx  \sum_{n}e^{-i\omega_{\rm d}n t}c_{n}\p{t}\ket{n,n\times \gd}
\label{eq:psioft}
\end{split}\end{equation}
for some time-dependent functions $c_{n}\p{t}$. As described above, these eigenstates lie along a line with slope $\gd$ in the spectrum shown in \cref{fig:spectrot}. This form can be observed in \cref{fig:prepoft} for a numerical simulation of a system with very weak driving.

In practice, the approximation of \cref{eq:appxdrlldE,eq:appxdrop} are insufficient to describe the dynamics. For any fixed $n/N$, the energy spacing between states shrinks exponentially with $N$ as we increase the size of the lattice, violating the assumption of \cref{eq:appxdrlldE}. For weak driving, the qualitative picture remains similar but the prepared state seen in \cref{eq:psioft} acquires a finite width in energy space, proportional to the driving strength. These features are seen in \cref{fig:stateprep}.

\subsubsection*{Observing many-body properties}

We've discussed above how to prepare the HCB system in a coherent-like state. This state has a defined kinetic energy per excitation, but it does not have a definite excitation number. We argue that this is not an impediment to measuring the many-body properties described above.

First, we note that for any measurements purely in the $\hat\gs^{z}$ basis we can effectively project the state into a definite $n$ sector by post-selection. This is useful for measuring, e.g., the correlation length shown in \cref{fig:rescorr}.

Second, we have observed in \cref{fig:resent} that the many-body properties that we are interested in behave similarly in different $n$ sectors of the spectrum. For these, we expect the state in  \cref{eq:psioft} to exhibit the same behavior as a function of its kinetic energy.

As such, preparing these coherent-like states may allow us to measure many-body properties of the spectrum by varying the detuning $\gd$. We verify this numerically in \cref{fig:stateprepmeas}, where we described a state prepared this way and measure its many-body properties. We find that the correlation length of the state, shown in \cref{fig:rescorr}, approximates very well the eigenmode correlation length for states in similar energy show in \cref{fig:prepmeasxi}. Similarly, the entanglement entropy measured as shown in \cref{fig:prepmeasent} exhibits the same behaviors we pointed out in \cref{fig:resent}.

\section*{Discussion}
We have offered here a roadmap for the realization of a quantum many-body simulator of the 2D Hard-Core Bose-Hubbard model using a superconducting circuit made up of transmon qubits. An experimental realization of this setup would allow the exploration of this analytically hard-to-solve model in regimes where it has not been realized before. In particular, we have shown how such a realization could access non-equilibrium states that exhibit many-body wavefunction behaviors such as a crossover from volume-law to area-law entanglement. As discussed throughout, the experimental parameters we consider in this article are within reach of current fabrication and control systems. The system we have proposed could be realized in the near term. 

In the body of this paper we've presented numerical results for a $4\times4$ HCB lattice, which can be diagonalized on a moderately powerful computer. However, the difficulty of this task grows exponentially, and a system of $6\times6$ or $7\times7$ sites is beyond numerical reach for any reasonable resource expenditure. An experimental realization would thus provide an example of quantum simulation beyond our theoretical and numerical abilities.

Beyond the model presented here, the paradigm of the QMBS can be used to explore a variety of other systems. Two immediate extensions of the model include changing the lattice topology or varying individual qubit frequencies to understand the role of disorder in this many-body system.
In the longer term, it would be interesting to explore other parts of the phase diagram in \cref{fig:phase}. In particular, a reliable and long-lived qubit with large anharmonicity would allow us to realize spin systems and explore their rich physics, including probing phase transitions and understanding spin liquids.

\section*{Methods}
\subsection*{Driving through a line coupled to multiple qubits \label{app:driving}}

Here, we give the derivation for the driving operator of \cref{eq:DriveOp}.

The system used, schematically shown in \cref{fig:implschem}, is described by the Hamiltonian
\begin{equation}
\hat H = \hat H_{\rm HCB} + \sum_{\ell}\hat H^{\rm L}_{\ell} + \sum_{\ell}\sum_{i\in \subS_{\ell}}\hat H^{\rm R}_{\ell,i},
\label{eq:drAll}
\end{equation}
where $\hat H_{\rm HCB}$, given in \cref{eq:HCB}, describes the qubits, $\hat H^{\rm L}_{\ell}$ the signal line $\ell$,
\begin{equation}
\hat H^{\rm L}_{\ell} = \intrm{d\nu}\nu \p{\hat L^{\ell\dagger}_{\nu}\hat L^{\ell}_{\nu} + \hat R^{\ell\dagger}_{\nu}\hat R^{\ell}_{\nu}}
\end{equation}
and $\hat H^{\ell}_{i}$ the resonator coupling line $\ell$ to qubit $i$,
\begin{equation}\begin{split}
 &\hat H^{\rm R}_{\ell,i}  = \p{\omega_{\rm q} + \Delta_{i}}\hat c_{i}\dg\hat c_{i} + g_{i}\p{\hat c_{i} + \hat c_{i}\dg}\p{\hat \gs^{+}_{i} + \hat \gs^{-}_{i}}
	\\ &  -i\tfrac{\sqrt{\gk_{i}}}{\sqrt{2}}\intrm{\tfrac{d\nu}{\sqrt{2\pi}}}\br{\p{e^{i\nu \tau_{i}}\hat R^{\ell\dagger}_{\nu} + e^{-i\nu \tau_{i}}\hat L^{\ell\dagger}_{\nu}}\hat c_{i} - \hc}.
\label{eq:HRi}
\end{split}\end{equation}
Here, $\ell$ sums over the different signal lines; for each line, $\hat R^{\ell}_{\nu}$ ($\hat L^{\ell}_{\nu}$) are the annihilation operators for its right (left) moving modes with energy $\nu$, and $\subS_{\ell}$ is the set of qubits coupled to it through resonators. For each resonator coupled to qubit $i$, $\hat c_{i}$ is the annihilation operator for a photon in the resonator, and $\Delta_{i}, g_{i},\gk_{i},\tau_{i}$ are that resonator's detuning, its coupling to the qubit $i$, its linewidth, and its distance from the termination of the signal line (divided by the speed of light), respectively. This setup is outlined in \cref{fig:implschem}.

Using standard input-output theory \cite{Gardiner2004}, the Heisenberg-Langevin equations of motion for the operators $\hat c_{i}$ are 
\begin{equation}\begin{split}
\dot{\hat c}_{i}\p{t} & = -\br{\half[\gk_{i}] + i\p{\omega_{\rm q} + \Delta_{i}}}\hat c_{i}\p{t}
	\\ & \qquad - \half[\sqrt{\gk_{i}}]\sum_{j\ne i}\sqrt{\gk_{j}}\hat c_{j}\p{t - \abs{\tau_{i}-\tau_{j}}}
	\\ & \qquad + \tfrac{\sqrt{\gk_{i}}}{\sqrt{2}}\p{\hat \xi^{\rm R}_{\ell}\p{t+\tau_{i}} + \hat\xi^{\rm L}_{\ell}\p{t-\tau_{i}}}
\end{split}\end{equation}
where $\hat \xi^{\rm L}_{\ell},\hat \xi^{\rm R}_{\ell}$ are Gaussian white noise operators describing the vacuum fluctuations of the left-moving and right-moving modes, respectively, on the line $\ell$ coupled to $i$.

If we drive the line at frequency $\omega_{\rm d}$, we have
\begin{equation}
\avg{\hat \xi^{\rm R}_{\ell}} \to \gO e^{-i\omega_{\rm d} t} \qquad \avg{\hat \xi^{\rm L}_{\ell}} \to -\gO e^{-i\omega_{\rm d}t}
\end{equation}
where $\gO$ is the driving field. We find, in steady state,
\begin{equation}\begin{split}
\avg{\hat c_{i}\p{t}} &
	=  -i\frac{\sqrt{2\gk_{i}}\gO \sin\p{\omega_{\rm d}\tau_{i}}}{\half[\gk_{i}] + i\p{\omega_{\rm q} + \Delta_{i} - \omega_{\rm d}}} e^{-i\omega_{\rm d}t}
\\ & \quad - \frac{1}{2}\sum_{j\ne i}\frac{\sqrt{\gk_{i}\gk_{j}}\avg{\hat c_{j}\p{t - \abs{\tau_{i}-\tau_{j}}}}}{\half[\gk_{i}] + i\p{\omega_{\rm q} + \Delta_{i} - \omega_{\rm d}}}.
\end{split}\end{equation}
In a dispersive readout scheme the linewidths of the cavities are narrow \cite{Blais2004},
\begin{equation}
\sqrt{\gk_{i}\gk_{j}}\ll\abs{\Delta_{i}}.
\end{equation}
If we then drive near the qubit frequency,
\begin{equation}
\abs{\omega_{\rm d} - \omega_{\rm q}}\ll \abs{\Delta_{i}},
\end{equation}
we can approximate
\begin{equation}\begin{split}
\avg{\hat c_{i}\p{t}} &\approx -\frac{\sqrt{2\gk_{i}}\gO}{\Delta_{i}} \sin\p{\omega_{\rm d}\tau_{i}} e^{-i\omega_{\rm d}t}.
\label{eq:avci}
\end{split}\end{equation}

Now, from \cref{eq:drAll,eq:HRi}, we have that the driving Hamiltonian can be described by
\begin{equation}
\hat H\to \hat H_{\rm HCB} + \sum_{\ell}\sum_{i\in \subS_{\ell}}g_{i}\p{\avg{\hat c_{i}} + \avg{\hat c_{i}\dg}}\p{\hat \gs^{+}_{i} + \hat \gs^{-}_{i}}.
\end{equation}
and for $\omega_{\rm d}\sim \omega_{\rm q}$, we can take the rotating wave approximation and combined with \cref{eq:avci} we find
\begin{equation}
\hat H \approx \hat H_{\rm HCB} + \sum_{\ell}\tilde g_{\ell}\p{e^{-i\omega_{\rm d}t} \hat \Sigma_{\ell}\dg  + e^{i\omega_{\rm d}t}\hat \Sigma_{\ell}},
\end{equation}
where
\begin{equation}
\hat \Sigma_{\ell} = \sum_{i\in \subS_{\ell}} \tilde \ga_{i}\hat \gs^{-}_{i},
\label{eq:DriveOpAppx}
\end{equation}
as in \cref{eq:DriveOp}, and 
\begin{align}
\tilde g_{\ell} & = \gO \sqrt{2K_{\ell}}, \label{eq:driveopdetailedg}
\\ \qquad \tilde \ga_{i} & = -\frac{\sqrt{\gk_{i}}g_{i}}{\sqrt{K_{\ell}}\Delta_{i}} \sin\p{\omega_{\rm d}\tau_{i}}, \label{eq:driveopdetailedga}
\\ K_{\ell} & = \sum_{i\in \subS_{\ell}}  \gk_{i}\tfrac{g^{2}_{i}}{\Delta_{i}} \sin^{2}\p{\omega_{\rm d}\tau_{i}}. \label{eq:driveopdetailedK}
\end{align}

\subsection*{Circuit analysis of the floating transmon qubit \label{app:floatingtransquant}}

\begin{figure}[t] 
   \centering
   
   \subfloat[\label{fig:circfl-qu}Floating transmon]{\includegraphics[scale=0.55]{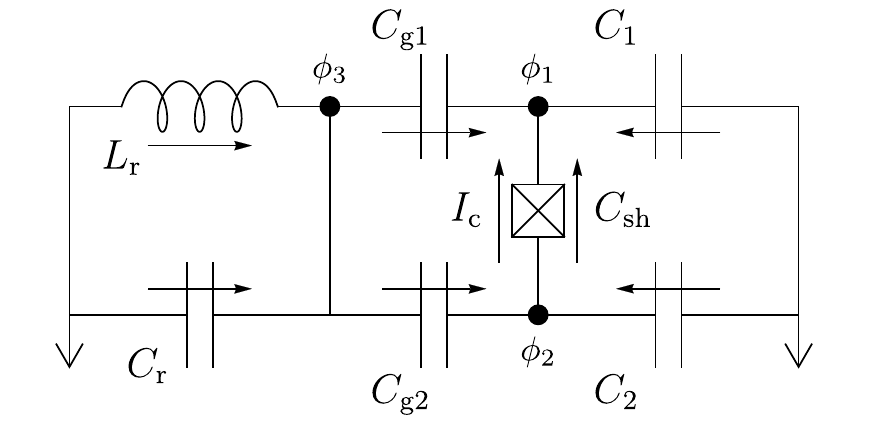}}
   \subfloat[\label{fig:circgr-qu}Grounded transmon]{\includegraphics[scale=0.55]{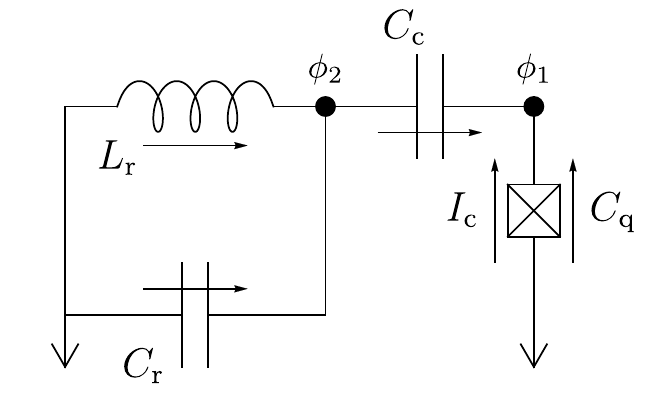}}

   \caption{\rev{\bf Parasitic capacitance in floating vs grounded transmon.} Circuit diagrams for a resonator coupled to \protect\subref{fig:circfl-qu} a floating transmon qubit and \protect\subref{fig:circgr-qu}  a grounded transmon. Independent nodes are labeled with their respective node phase $\phi_i=2\pi\Phi_i/\Phi_0$, relating to node fluxes $\Phi_i$.}
   \label{fig:flgr-quant}
\end{figure}

We review the Hamiltonian of a floating transmon qubit coupled to a harmonic oscillator mode, depicted in \cref{fig:circfl-qu}. This allows us to extract effective values for the qubit capacitance $C_{\mathrm{q}}$ and the coupling capacitance $C_{\mathrm{c}}$ comparable to those of a grounded transmon qubit,  shown in \cref{fig:circgr-qu}. 
In the Results section, we utilize this result to compare unwanted crosstalk in an architecture with floating transmon qubits versus an architecture that makes use of grounded  transmons.

Following the node flux representation described in Ref.~\onlinecite{Vool2017} we can write down the Lagrangian for the circuit in \cref{fig:circfl-qu} as

\begin{equation}
\mathcal{L} = \mathcal T - \frac{\Phi_3 ^2}{2L_{\mathrm{r}}} + E_{\mathrm{J}}\cos(\phi_1-\phi_2),
\end{equation}
\begin{equation} \begin{split}
\mathcal{T}& =\frac{C_1}2 \dot\Phi _1 ^2 +\frac{C_2}2 \dot\Phi _2 +\frac{C_{\mathrm{sh}}}2 (\dot\Phi _1-\dot\Phi _2) ^2 
	\\ &\quad +\frac{C_{\mathrm{r}}}2\dot\Phi _3 ^2 +\frac{C_{\mathrm{g1}}}2 (\dot\Phi _1-\dot\Phi _3) ^2 +\frac{C_{\mathrm{g2}}}2 (\dot\Phi _2-\dot\Phi _3) ^2,
\label{eq:fltrT}
\end{split}\end{equation}
where $\Phi_i$ are node fluxes and $\phi_i=2\pi\Phi_i/\Phi_0$ node phases, with $\Phi_0$ the magnetic flux quantum. $E_{\mathrm{J}}=\Phi_0I_{\mathrm{c}}/2\pi$ is the Josephson energy of the Josephson junction with critical current $I_{\mathrm{c}}$. The kinetic part of the Lagrangian can also be written as
\begin{equation}
\mathcal T = \frac{1}{2} \dot{\vec \Phi}^{\rm T} \cdot \check C \cdot \dot{\vec \Phi}
\end{equation}
where $\vec\Phi=\mat{\Phi_{1} & \Phi_{2}& \Phi_{3}}^{\mathrm{T}}$ and $\check C$ the capacitance matrix defined by \cref{eq:fltrT}.

In order to recover the relevant transmon degree of freedom, we perform a variable transformation in the transmon subspace to `plus-minus' variables ${\Phi_\pm=\Phi_1\pm\Phi_2}$. With the transformation matrix
\begin{equation}
S=\left(\begin{array}{ccc}1&1&0\\1&-1&0\\0&0&1\end{array}\right)
\end{equation}
we can rewrite the capacitive part of the Lagrangian as
\begin{equation}\begin{split}
\mathcal{T} & =\frac 1 2 \p{S \dot{\vec\Phi}}^{\mathrm{T}}\cdot {S^{-1} \check C  S^{-1}}\cdot {S \dot{\vec\Phi}}
\equiv \frac{1}{2} {\dot{\vec \Phi}^{\prime \rm T}}\cdot \check{\mathcal C} \cdot \dot{\vec \Phi}\pr.
\end{split}\end{equation}
Here $\vec\Phi\pr = S\cdot \vec \Phi=\mat{\Phi_{+} & \Phi_{-}& \Phi_{3}}^{T}$, and the capacitance matrix in the transformed basis becomes $\check{\mathcal C}=S^{-1}\check C S^{-1}$. 

A Legendre transformation yields the circuit Hamiltonian
\begin{equation}
\mathcal H=\frac 1 2 \vec q^{\prime \rm T}\cdot \check {\mathcal C}^{-1}\cdot {\vec q}\pr + \frac{\Phi_3 ^2}{2L_{\mathrm{r}}} - E_{\mathrm{J}}\cos\phi_{-},
\end{equation}
where ${\vec q}\pr = \mat{q_{+} & q_{-} & q_{3}}$.
Since the `$+$'-mode of the transmon does not have an inductive term in the Hamiltonian, its frequency is not relevant for qubit operation. Conversely, the Josephson energy of the transmon enters via the `$-$'-mode. We can therefore trace over the $q_{+}$ degree of freedom, to find
\begin{equation}
\mathcal H\to \frac 1 2 \mat{q_{-}\\ q_{3}}\cdot \br{\Tr_{+}\check {\mathcal C}^{-1}}\cdot \mat{q_{-}\\ q_{3}}^{\rm T} + \frac{\Phi_3 ^2}{2L_{\mathrm{r}}} - E_{\mathrm{J}}\cos\phi_{-},
\end{equation}
where
\begin{equation}
\Tr_{+}\check {\mathcal C}^{-1} = \mat{\br{\check {\mathcal C}^{-1}}_{-,-} & \br{\check {\mathcal C}^{-1}}_{-,3} \\ \br{\check {\mathcal C}^{-1}}_{3,-} & \br{\check {\mathcal C}^{-1}}_{3,3}}
\label{fig:quantHtraced}
\end{equation}
is the matrix $\check {\mathcal C}^{-1}$ with the column and row corresponding to the $+$ mode removed.

The effective Hamiltonian of \cref{fig:quantHtraced} has the same form as the Hamiltonian resulting from analysis of the circuit \cref{fig:circgr-qu}, with the substitutions ${\Phi_{1}\to \Phi_{-}}, {q_{1}\to q_{+}}$. We can then find the effective parameters of the reduced circuit by identifying
\begin{equation}\begin{split}
 \br{\Tr_{+}\check {\mathcal C}^{-1}}^{-1}&  \equiv \check C_{\rm eff} = \mat{C_{\rm q} + C_{\rm c} & - C_{\rm c} \\ - C_{\rm c} & C_{\rm r} + C_{\rm c}}.
\end{split}\end{equation}

We therefore find the effective transmon capacitance, including coupling capacitances to the resonator, from the diagonal entry,
\begin{equation}\begin{split}
C_{\rm q,eff}&  = C_{\rm q} + C_{\rm c} = 
\\ &\quad C_{\mathrm{sh}}+\left(\frac1{C_1+C_{\mathrm{g}1}}+\frac1{C_2+C_{\mathrm{g}2}}\right)^{-1},
\end{split}\end{equation}
and from the off-diagonal entries we can extract the effective coupling capacitance between the floating transmon qubit and the resonator
\begin{equation}
C_{\mathrm{c}}=\frac{C_{\mathrm{g}1}C_2 -C_{\mathrm{g}2}C_1}{C_1+C_2+C_{\mathrm{g}1}+C_{\mathrm{g}2}}.
\end{equation}

Applied to the circuit in \cref{fig:circfl}, we find the parasitic coupling between the floating transmon qubit and the resonator
\begin{equation}
C_{\rm eff}^{\rm \p{f}}=\frac{C_{\rm G}\p{C_{\rm P}-C_{\rm P}\pr}}{2C_{\rm G}+C_{\rm P}+C_{\rm P}\pr}
\end{equation}
taking $C_{\rm g1}=C_{\rm g2}= C_{\rm G}$.

\section*{Data availability}

The results of the simulations generated during the study are available from the corresponding author on reasonable request and with the approval of our US Government sponsor.

\section*{Code availability}

The simulation code generated during the study is available from the corresponding author on reasonable request and with the approval of our US Government sponsor.

\section*{Author Information}
\subsection*{Contributions}
YY, CT, and WDO devised to the initial concept.
YY performed the analysis and numerical simulations of the HCB and coherent-like state preparation.
JB performed the analysis related to the floating transmon implementation.
YY and JB and wrote the paper, and all the authors contributed to the discussions.

\section*{Ethics declarations}
\subsection*{Competing interests}
The authors declare that there are no competing interests.

\bibliographystyle{PRlike}

\bibliography{ManyQubitProposal}

\end{document}